\documentclass{article}

\usepackage{amsmath}
\usepackage{arxiv}

\usepackage[utf8]{inputenc} 
\usepackage[T1]{fontenc}    
\usepackage{hyperref}       
\usepackage{url}            
\usepackage{booktabs}       
\usepackage{amsfonts}       
\usepackage{nicefrac}       
\usepackage{microtype}      
\usepackage{lipsum}
\usepackage{graphicx}
\graphicspath{ {./images/} }

\title{An Evaluation of Generative Pre-Training Model-based Therapy Chatbot for Caregivers}

\author{
  Lu Wang \\
  College of Computing \& Informatics\\
  Drexel University\\
  Philadelphia, PA 19104 \\
  \texttt{lw823@drexel.edu} 
  \And
  Munif Ishad Mujib \\
  College of Computing \& Informatics\\
  Drexel University\\
  Philadelphia, PA 19104 \\
  \texttt{mim52@drexel.edu} 
  \And
  Jake Williams \\
  College of Computing \& Informatics\\
  Drexel University\\
  Philadelphia, PA 19104 \\
  \texttt{jw3477@drexel.edu} 
  \And
  George Demiris \\
  School of Nursing, Perelman School of Medicine\\
  University of Pennsylvania\\
  Philadelphia, PA 19104 \\
  \texttt{gdemiris@nursing.upenn.edu} 
  \And
  Jina Huh-Yoo \\
  College of Computing \& Informatics\\
  Drexel University\\
  Philadelphia, PA 19104 \\
  \texttt{jh3767@drexel.edu} 
}

\begin{document}
\maketitle
\begin{abstract}
With the advent of off-the-shelf intelligent home products and broader internet adoption, researchers increasingly explore smart computing applications that provide easier access to health and wellness resources. AI-based systems like chatbots have the potential to provide services that could provide mental health support. However, existing therapy chatbots are often retrieval-based, requiring users to respond with a constrained set of answers, which may not be appropriate given that such pre-determined inquiries may not reflect each patient's unique circumstances. Generative-based approaches, such as the OpenAI GPT models, could allow for more dynamic conversations in therapy chatbot contexts than previous approaches. To investigate the generative-based model's potential in therapy chatbot contexts, we built a chatbot using the GPT-2 model. We fine-tuned it with 306 therapy session transcripts between family caregivers of individuals with dementia and therapists conducting Problem Solving Therapy. We then evaluated the model's pre-trained and the fine-tuned model in terms of basic qualities using three meta-information measurements: the proportion of non-word outputs, the length of response, and sentiment components. Results showed that: (1) the fine-tuned model created more non-word outputs than the pre-trained model; (2) the fine-tuned model generated outputs whose length was more similar to that of the therapists compared to the pre-trained model; (3) both the pre-trained model and fine-tuned model were likely to generate more negative and fewer positive outputs than the therapists. We discuss potential reasons for the problem, the implications, and solutions for developing therapy chatbots and call for investigations of the AI-based system application.\end{abstract}


\section{Introduction}
Acute mental health provider shortages alarmed the Association of American Medical Colleges in 2016 \cite{butryn2017shortage}. Eighteen percent of the counties in the U.S. had reported a lack of providers, such as psychologists, social workers, advanced practitioners, therapists, and counselors \cite{thomas2009county}. Nearly every county (e.g., 96\% of $3,140$ counties in the U.S.) needed more psychiatrists \cite{thomas2009county}. Rural counties and low-income counties have higher levels of need concerning healthcare access \cite{thomas2009county}. This issue is a global problem to this day. In developed countries, 100,000 people approximately share 6.6 psychiatrists, while, in lower-income countries, 100,000 people approximately share as few as 0.1 psychiatrists \cite{oladeji2016brain}. An estimated additional 1.18 million mental health workers are needed in low- and middle-income countries for basic mental health interventions \cite{oladeji2016brain}. Developing cost-effective and feasible mental health support systems to mitigate these shortages will be critical. 

Chatbots are digital tools that interact with users in natural language, often for the goal of helping a user complete a task, \cite{laranjo2018conversational, vaidyam2019chatbots}. Chatbots, in the form of voice agents with speech functions became widely available to the public through off-the-shelf products, such as Apple’s Siri, Microsoft’s Cortana, and Amazon’s Alexa \cite{belfin2019graph}. Having been adopted in multiple fields, including business \cite{adam2020ai}, governance \cite{androutsopoulou2019transforming}, and education \cite{georgescu2018chatbots}, the potential of chatbots is also presented in mental health, where chatbots are labeled as “the future of therapy” \cite{vaidyam2019chatbots} and are believed to increase the opportunities for individuals to receive therapeutic or emotional support services \cite{moore2019bot}.  In 2019, Facebook Messenger had more than 300,000 chatbots and many were classified as related to healthcare and wellbeing \cite{zhang2020artificial}. Several studies have proven their efficacy: therapy services delivered through chatbots were as effective as face-to-face therapy in diagnosing and treating anxiety and depression symptoms \cite{fulmer2018using, ho2018psychological, inkster2018empathy, roniotis2017detecting}. Therapy chatbots provide economic and easily accessible services through multiple forms (e.g., text, chat app, social media) and personalized, immediate support available 24/7 \cite{palanica2019physicians, zhang2020artificial}. Especially when integrated into the health and ambient assistive technologies (e.g., “ambiently used sensor-based information and communication technologies, aiming at contributing to a person’s health and health care as well as to her or his quality of life” \cite{haux2016health}), therapy chatbots will bring increasing possibility in personal healthcare including prevention, intervention, and therapy by coordinating with other sensors and ambient, smart health solutions as home settings will become a critical place to perform clinical care, in addition to hospitals and clinics \cite{haux2016health}.

Therapy chatbots need to address users’ emotional needs, unlike other social chatbots that offer information only \cite{sharma2018digital}. Furthermore, therapy chatbots need to be distinguished from generic, social chatbots in that they involve persuasion, which is a conceptualized more complex task than those of social chatbots that engage users in general conversations (e.g., talking about movies or weather) \cite{zhang2020artificial}. Therapy contexts would require the chatbots to generate responses to complicated questions and allow users to lead the conversation, which is challenging but could be achieved by increasingly advancing artificial intelligence techniques developed in recent years \cite{abd2019overview, fulmer2018using, gao2018neural}. In comparison, social chatbots may be evaluated with standard metrics, such as BLEU, METEOR, and ROUGE-N family \cite{chen2017survey, lin2004rouge}, evaluating the quality of responses, such as their continuity, engagement, compassion, and coherence, is critical \cite{chen2017survey, liu2016not}. However, very few studies have evaluated the speech quality of chatbots \cite{chen2017survey, liu2016not}, however, the need has been documented in depth \cite{chaves2020should, zhang2020artificial}. And few studies have discussed the possible negative consequences of applying undesirable therapy chatbots, especially the ethical problems \cite{fleischmann2019good}, even though many therapy chatbots have been applied with some constraints in delivering therapy supports.

Numerous platforms are being released using advancing machine learning techniques, such as RNN (Recurrent Neural Network), LSTM (Long Short Term Memory), and Seq2Seq model (Encoder-Decoder Sequence to Sequence Models) \cite{dsouza2019chat}. These platforms bring opportunities for building effective therapy chatbots at a low cost. Most therapy chatbots, however, force users to respond among pre-determined options. Such forms of communication do not suit therapy contexts in which patients need to be open, honest, and expressive. With an interest in building a therapy chatbot that allows more freedom of conversation for therapy, we investigate basic qualities of a state-of-the-art technique called Generative Pre-Training Model-2 (GPT-2) \cite{radford2019language, OpenAIwebsite} for therapy chatbot contexts. Based on our results, we discuss the implications of designing and building therapy chatbots contributing to the field's discussion around human-centered AI design.

\section{Related Work}
In this section, we walk through a few seminal approaches to building and evaluating therapy chatbots.

\subsection{Building Therapy Chatbot Models}

Largely two main conversational approaches make up the models of building chatbots: retrieval-based and generative-based \cite{dsouza2019chat, mudikanwi2011student}. The key distinction between the two lies in that retrieval-based chatbots find matched responses from the database of manually created utterances of conversations. In contrast, the generative-based chatbots auto-generate responses via machine learning algorithms. 

To date, most therapy chatbots apply a retrieval-based approach \cite{abd2019overview, laranjo2018conversational}. Based on response matching, retrieval-based chatbots rely on dialogue management frameworks to track conversation states \cite{chen2017survey} and decide future actions \cite{swartout2013virtual}. Most therapy chatbots have used hand-crafted dialogue management frameworks \cite{thomson2013statistical} of finite states \cite{sutton1998universal} and frame-based, also known as form-filling, frameworks \cite{goddeau1996form}. For the framework of the finite state, the dialogue is constrained to a sequence of pre-determined steps or states \cite{laranjo2018conversational}, and users are required to select responses for single-turn conversations \cite{chen2017survey}. This goes well with straightforward and well-structured tasks but will fail if users need to take initiative in the conversation \cite{laranjo2018conversational}. For a frame-based or form-filling framework, the flow of the dialogue is not pre-determined \cite{laranjo2018conversational} but proceeds according to the pre-specified action for each pre-defined set of known concepts called slots \cite{thomson2013statistical}. This kind of framework is usually used in information-seeking conversations \cite{bohus2009ravenclaw}, where users seek information according to a set of constraints. An example of this is where users provide information to fill slots, such as the departure to and arrival in a city to search a route. However, this framework sometimes struggles to adapt to other kinds of conversations \cite{thomson2013statistical} and often causes users to provide more information than needed due to non-predetermined dialogue flow \cite{laranjo2018conversational}. Several popular techniques to realize the dialogue management frameworks are Artificial Intelligence Markup Language (AIML) and ChatScript etc. \cite{dsouza2019chat}. AIML, firstly adopted by ALICE (Artificial Linguistic Internet Computer Entity), is an XML-compliant language that allows for efficient pattern matches in a tree structure for retrieving responses. Seminal therapy chatbots reported in the literature—VICA, a virtual agent equipped with voice-communication \cite{sakurai2019vica}, a conversational agent for alcohol misuse intervention \cite{elmasri2016conversational}, and a counseling agent in the IT industry \cite{shinozaki2015context}—all applied AIML to build the chatbot. Vivibot \cite{greer2019use}, Woebot \cite{fitzpatrick2017delivering}, and a virtual agent for post-traumatic stress disorder \cite{tielman2017therapy} also applied decision tree structures. An embodied conversational agent for education \cite{sebastian2017changing} applied the option-choice format to allow user replies. However, the retrieval-based design allows chatbots to reply with more coherent answers than generative-based design \cite{dsouza2019chat}, it restrains free conversations \cite{klopfenstein2017rise, laranjo2018conversational, zhang2020artificial} due to pre-created outputs \cite{mudikanwi2011student}. It is insufficient for multi-linear conversations due to the decision tree mechanism \cite{dsouza2019chat}. Additionally, it will fail the task if users’ inputs do not match any database \cite{dsouza2019chat}, making it difficult to improve usability. 

Alternatively, generative-based chatbots allow for conversational flexibility. This model applies machine learning techniques to train the chatbots to learn and generate responses based on a large amount of training data \cite{dsouza2019chat}. Popular artificial intelligence techniques are RNN, LSTM, and Seq2Seq model \cite{dsouza2019chat, trivedi2019chatbot}. Few studies have tried to apply a g generative-based approach to build therapy chatbots. Among the generative-based models, the state-of-the-art models include Bidirectional Encoder Representations from Transformers (BERT) \cite{devlin2018bert} and the OpenAI Generative Pre-Training-2 Model (GPT-2) \cite{radford2019language}, which has been expanded to a third-generation, autoregressive language model (GPT-3) \cite{brown2020language}. These models are open-sourced, efficient to model training, and tailorable for task-oriented dialog generation \cite{qiu2020pre, zhang2020artificial}. The OpenAI GPT-2 as a generative unsupervised pre-trained model was released in 2019 and trained on a large unlabeled training corpus, which can reduce manual annotation costs, avoid training a new model from scratch and allow for deep language models \cite{OpenAIwebsite, qiu2020pre}. Tests showed the model achieved state-of-the-art performance on language tasks like question answering, reading comprehension, summarization, and translation \cite{OpenAIwebsite, radford2019language}. The chatbot can also be fine-tuned with different domain data for unique purposes for its target users \cite{lee2020patent, vig2019multiscale}. However, problems exist, like users having difficulty understanding and model generating errors that violate common sense \cite{zhang2020artificial}. One general solution is to incorporate pre-trained models to facilitate conversations in specialized domains by fine-tuning with domain’s datasets \cite{OpenAIwebsite}. 

\subsection{Evaluation of Therapy Chatbots}

Conducting evaluations on chatbots \cite{laranjo2018conversational} range from technical performance, user experience and to speech quality. 

\textbf{Technical performance.} Retrieval-based chatbots are evaluated based on the rate of successful task completion and recognition accuracy of speech \cite{laranjo2018conversational}. Typical measurements include accuracy, which refers to the percentage of label matched, and Precision, Recall, and F-measure, which are based on relevance \cite{dsouza2019chat}. In contrast, generative-based chatbots are evaluated using Word Similarity Metrics such as BLEU, METEOR, and ROUGE-N family for their technical performance \cite{chen2017survey, lin2004rouge}. Furthermore, datasets such as the corpus of CNN/Daily Mail dataset \cite{nallapati2016abstractive}, the Gigaword corpus \cite{napoles2012annotated}, the 2004 Document Understanding Conference dataset \cite{harman2004effects}, arXiv \cite{scharpf2020classification}, and PubMed \cite{dynomant2019doc2vec} are provided and widely used to evaluate the generated responses of chatbots, allowing researchers to compare models’ performances based on the Word Similarity Metrics. Although these metrics are frequently used, researchers found that they are either weak or have no correlation to human judgments even though they can serve as measurements to distinguish state-of-the-art models from baselines \cite{liu2016not}. One promising method is to employ an approach to distinguish models’ outputs from those produced by humans \cite{chen2017survey}. 

\textbf{User experience.} Research in therapy chatbots applied user research to evaluate user experience , including measuring users’ trust and comfortability \cite{sakurai2019vica}, emotional states \cite{fitzpatrick2017delivering, fulmer2018using, greer2019use}, overall satisfaction \cite{elmasri2016conversational}, and acceptability and usability outcomes \cite{fitzpatrick2017delivering,laranjo2018conversational, tielman2017therapy}. Several researchers used the Positive and Negative Affect Schedule (PANAS) to test emotional states \cite{fitzpatrick2017delivering, fulmer2018using, joerin2019psychological}. However, user research is often costly or limited to small samples \cite{fitzpatrick2017delivering, greer2019use}. 

\textbf{Speech quality.} Speech quality \cite{moller2000new} examines the gap between the user’s perception and expectation during the conversation based on the context. Unlike technical performance evaluations, which have focused on the general performance of chatbots in language generation, speech quality measures the effectiveness of conversation delivered by the chatbots in the specific application contexts. Zhang et al. discussed measuring the speech quality of chatbots through either subjective evaluation from the user’s perspectives (e.g., coherence, naturalness, and fluency) or objective evaluation (e.g., linguistic analyses of contents, lengths of conversations, and amounts of information exchanged) \cite{zhang2020artificial}. Objective evaluation, including meta-information of the conversation (e.g., utterance length, turn-taking, words used, etc.), is especially suitable for the generative-based approach. The responses are auto-generated by the chatbots whose quality can not be guaranteed, unlike human moderated responses in the retrieval-based approach. Previous therapy chatbot research used similar evaluations to measure speech performance, such as the average number of interactions in a single dialogue session called Conversation-turns Per Session (CPS) \cite{sakurai2019vica, shinozaki2015context,shum2018eliza}. 

The OpenAI GPT-2 model has shown that it reached its benchmark in terms of its technical performance \cite{OpenAIwebsite, radford2019language}. However, such performance evaluations are not enough to explain the requirements needed for sensitive contexts, such as the safety and credibility that users experience. Assessing user experience requires putting human subjects at risk by exposing them to untested therapy chatbots. Given that this is the first step into evaluating a therapy chatbot using the generative-based model, we begin with assessing the basic qualities of speech and conversation measured through the meta-information of the chatbot responses. 

\section{Methods}

The pre-trained model refers to the released factory model without additional fine-tuning with the training data of an application area. The fine-tuned model is tuned by a domain-specific dataset based on the pre-trained model with an application goal. Our goal was to investigate how these pre-trained and fine-tuned models of the OpenAI GPT-2 perform as therapy chatbots. As a preliminary step into this long journey, we first focused on whether the chatbots respond with basic key conditions associated with speech quality that can be measured using meta-information on the words, length of words, and sentiments in chatbot's responses: 
\begin{itemize}
\item {RQ1}: How do chatbots with pre-trained and fine-tuned models perform in generating understandable responses?
\item{RQ2}: How do pre-trained and fine-tuned models perform in adjusting the information load with users’ inputs when compared to the therapists?
\item{RQ3}: What are the sentiments of the pre-trained and fine-tuned models compared to that of the therapists?
\end{itemize}

Below, we walk through the following: (1) generative-based model and the dataset we used to fine-tune and test the models and (2) the background of how we evaluated RQ2 and RQ3 in terms of adjusting the information load and sentiments used for therapist-patient interaction.

\subsection{Dataset and Fine-tuning}

Due to concerns with malicious applications, the OpenAI applied a staged release plan and shared four GPT-2 model sizes: 117, 345, 762, and 1542 million parameters. They are respectively called the 117M model, 345M model, 762M model, and 1.5GB model. We applied the Open AI GPT-2 345M model. It was the largest model available when we initiated the experiment, which was in September of 2019. As of December of 2020 (when we are writing this draft after completing the analysis), the 1.5GB model has been the most updated version so far. Since the OpenAI GPT-2 outperforms other language models trained on specific domains without fine-tuning \cite{OpenAIwebsite,radford2019language} and as an open-sourced pre-trained model, which was trained by over 8 million human filtered documents for a total of 40 GB of text, it can reduce manual annotation costs, time, and computational resources. We wanted to investigate how well the pre-trained model performs in a therapist-patient dialogue context. Meanwhile, we also fine-tuned the model to see whether fine-tuning with training data can bring better results than the pre-trained model. It is a potential method to reduce problems like topic changes and the model generating errors that violate common sense \cite{OpenAIwebsite}.

Although larger models might result in better outcomes with research goals\cite{lee2020patent}, some research has compared outcomes from different model sizes and demonstrated that the results are consistent between different model sizes \cite{xu2019neural}. GPT-3 is another updated model that has attempted to remove the fine-tuning process and build a general language model. However, researchers found that GPT-3 did not yield satisfying performance because it generated off-topic, confusing answers and had ethical issues, such as cultural biases in its responses \cite{floridi2020gpt}. The OpenAI has not open-sourced the model’s code yet, and the only API is available for the model’s capacity testing \cite{mcguffie2020radicalization}. For our goal, because GPT-3 does not allow fine-tuning, our approach of evaluating GPT-2 instead of GPT-3 remains the most updated approach to testing generative-based models that allow for fine-tuning, especially because of our domain context of the therapist-patient dialogue context.

We had access to 306 transcribed therapy sessions between 152 family caregivers of individuals with dementia and their respective mental health therapists. Among them, 59 transcriptions were excluded because they were transcripts unrelated to the main therapy sessions (e.g., the closing session of therapy where therapists reflect the whole process and say farewell to patients). Then, duplicate sessions, of which there were 8, were excluded, resulting in 239 sessions. This process follows the common practice in the machine learning study of splitting datasets by train/test \cite{boyd2019deep}, We then fine-tuned the model with 75\% of the 239 remaining sessions, i.e., 179 sessions from 123 caregivers. The remaining 25\% of the sessions were used for evaluation, i.e., 60 sessions from 52 caregivers, which consisted of 9261 patient-therapist response pairs. 

\subsection{Evaluation: Non-word outputs, response length, sentiment analyses}
To answer the three research questions, we used three measurements: the proportion of non-word outputs (RQ1), the length of response (number of words in a response) (RQ2), and sentiment components (RQ3). 

\subsubsection{Measurement for RQ1: The proportion of non-word outputs}
For the proportion of non-word outputs, when the model’s responses did not contain any English words (e.g., punctuations only like “????????”), we considered them as non-word outputs that failed the initial step to being evaluated for the model’s speech quality. We conducted a two-proportion test \cite{newcombe1998interval, newcombe1998two} to test the proportions of non-word outputs of the pre-trained versus fine-tuned models.

\subsubsection{Measurement for RQ2: Length of response (number of words)}
To guarantee successful communication, speakers and listeners should collaborate to allow the conversation to continue \cite{olmstead2020role}. According to the Conversation theory \cite{pask1976conversation} and the multi-action cascade model (MACM) of conversation \cite{gunaratne2019multi}, conversation participants can act in three ways to allow the dialogue to continue: (1) initiate a new conversation, (2) contribute to an existing conversation, and (3) acknowledge shared information (e.g., “I see” “That’s great!”) \cite{heritage2017conversation}. 

For conversations to proceed, two people who are conversing with one another overall maintain the balance of information, specifically regarding the length of responses \cite{olmstead2020role}. Lower information load cannot intrigue the other person to contribute to the conversation due to limited information to prompt conversations. A higher information load can make it harder for individuals to digest the information right away. Information overload refers to information load that is greater than the receiver’s information processing capacity \cite{roetzel2019information, streufert1965conceptual}. An ideal information load is neither too low nor too high for the receiver’s capacity, including their characteristics (such as serial processing ability or limited short-term memory) or limited task-related equipment (such as time or budget) \cite{roetzel2019information}. If therapy chatbots provide an unsuitable information load to users, dissatisfying or negative outcomes will occur. Hence, we evaluated the length of each output to assess whether the models responded with longer or shorter utterances compared to the therapists’ responses. 

Given the factors discussed above, the length of responses calculated as a total number of words per response is used as an indicator of the amount of information shared for each response  \cite{calvo2014finding}. Then we conducted a {\itshape One-way Repeated ANOVA} \cite{kim2015statistical} to test whether there was any significant difference in the response lengths among the three outputs from the pre-trained model, the fine-tuned model, and the therapists. If the {\itshape One-way Repeated ANOVA} result indicated there was a significant difference among the three outputs, we performed {\itshape Tukey’s HSD} test \cite{abdi2010tukey} for pairwise comparisons.

\subsubsection{Measurement for RQ3: Sentiment analysis}
\begin{figure}[htbp]
  \centering
  \includegraphics[width=10cm]{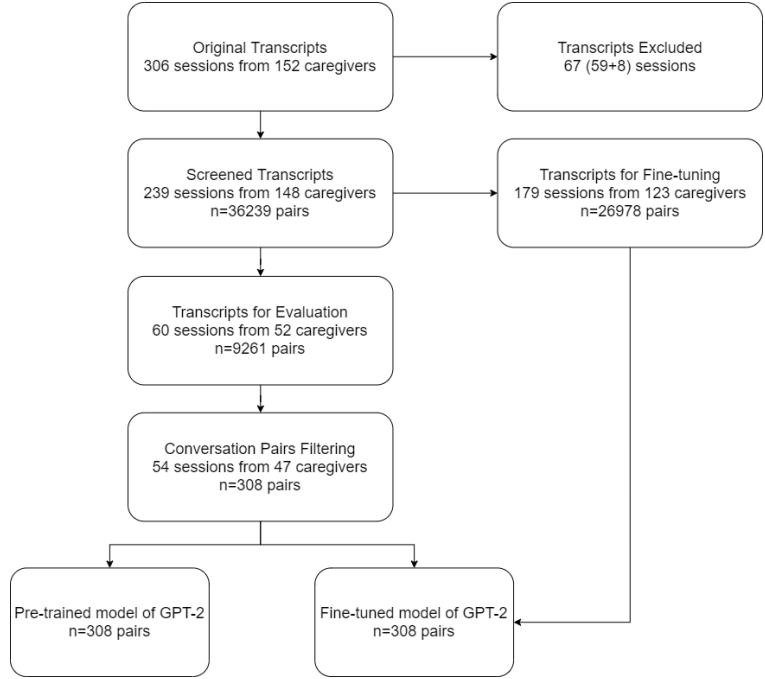}
  \caption{Diagram showing the process of filtering conversational pairs to evaluate and select the training dataset for the fine-tuned model.}
  \label{fig:Prisma}
\end{figure}

The therapists in the transcript data used Problem-Solving Therapy, an intervention for managing depression and anxiety \cite{zhang2018effectiveness}. In this technique, positive reinforcement is one of the fundamental key components in establishing a therapeutic alliance \cite{nezu2006problem}. To evaluate the level of positive reinforcement the models perform, we created a keyword list that would identify therapists’ original conversation pairs that included positive reinforcement among the 60 sessions set aside as evaluation data. In generating this keywords list, we used the SocialSent Casual ConversationA lexicon \cite{hamilton2016inducing}, a lexicon known to effectively convey sentiments that may differ according to the context of the conversation. We selected the keywords from the lexicon within 2 standard deviations of the mean according to the SocialSent Casual Conversation’s positive sentiment scores. Of these 4,621 keywords, 143 keywords appeared in the 60 sessions and 9261 conversation pairs. Two authors of this study conducted a manual annotation of randomly selected 100 conversation pairs to determine whether the keyword included in the conversation pair was relevant to therapists positively reinforcing the patients’ responses. The Cohen's Kappa for inter-rater reliability for this task was 0.62. For disagreements on the inclusion of keywords, the authors discussed the differences and agreed. The overall team then discussed the final keywords to include, which were 35 keywords in number. The resulting keywords included “Good,” “Yeah,” or “Nice.” If the therapist's utterance included at least one of these keywords, we selected the conversation pair for evaluation. This process resulted in a total of 308 conversation pairs, covering 54 sessions from 47 caregivers. We extracted the patient’s utterances as an input to the pre-trained and fine-tuned models to generate response outputs from the two models (See Fig. \ref{fig:Prisma}). 
We then compared the two responses from the pre-trained and fine-tuned models against the therapists’ original responses to evaluate their comparative sentiments. To do this, we used SÉANCE version 1.05 \cite{crossley2017sentiment} to calculate the two models’ outputs’ expanded sentiment scores on positivity. It is a freely available sentiment analysis tool, which contains an extensive database of dictionaries of words. Unlike most sentiment analysis tools, this tool integrates multiple sentiment analysis tools’ dictionaries, including the Harvard IV-4 dictionary lists used by the General Inquirer \cite{stone1966general}, the Lasswell dictionary lists \cite{lasswell1969value}, the Affective Norms for English Words \cite{bradley1999affective}, Hu–Liu polarity lists \cite{hu2004mining}, Geneva Affect Label Coder \cite{scherer2005emotions}, EmoLex \cite{mohammad2010emotions,mohammad2013crowdsourcing}, SenticNet \cite{cambria2012senticnet}, and the Valence Aware Dictionary for Sentiment Reasoning \cite{hutto2014vader}. SÉANCE generated 20 weighted component scores from the indices of these databases through principal component analysis. We chose 10 component scores relevant to positive reinforcement. These components were “Negative Adjectives,” “Positive Adjectives,” “Joy,” “Affect for Friends and Family,” “Fear and Disgust,” “Positive Nouns,” “Respect,” “Trust Verbs,” “Failure,” and “Positive Verbs.” We disregarded the remaining 10 weighted component scores generated by SÉANCE because they were not applicable in the context of this study. 

\section{Results}
\subsection{RQ1 findings: The proportion of non-word outputs}
The Fine-tuned model performed worse in generating more outputs that were not English words compared to the pre-trained model. The proportion of non-word outputs of the pre-trained model versus the fine-tuned model was 5.8\% (18 out of 308 conversation pairs) and 40.6\% (125 out of 308 conversation pairs). The two-proportion test \cite{newcombe1998interval,newcombe1998two} showed a significant difference between these two proportions: the 96\% {\itshape confidence interval} is [0.281, 0.408] and the {\itshape sample estimates} were 5.5\% and 39.9\% respectively. Examples of non-word outputs included: “????????”, “Â” and “\_\_\_\_.” Examples of remained outputs included: “I see why he would want to keep doing this,” “Wow! And these are things that you've sung with her before,” and “It went really well.”

\subsection{RQ2 findings: Length of response (number of words)}
\begin{figure}[h]
  \centering
  \includegraphics[width=10cm]{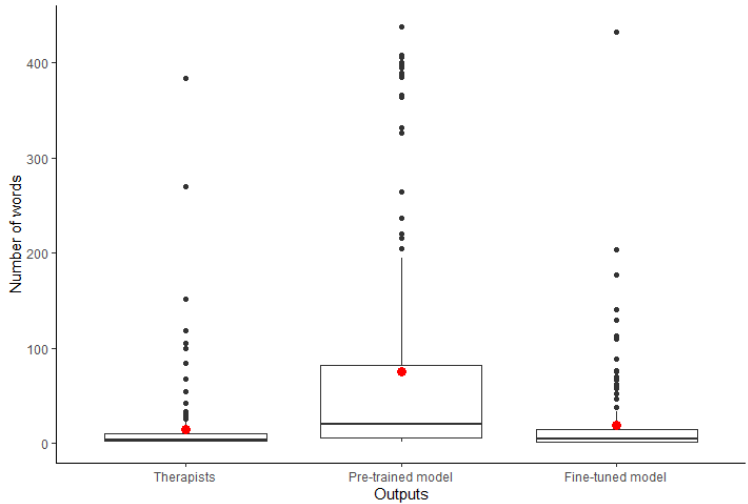}
  \caption{Boxplot of the number of words for each response of three outputs (The red points are mean values for each). Therapists’ responses and the fine-tuned model responses are similar in length, while the pre-trained model had a significantly higher number of words compared to the therapists’ responses in the transcripts.}
  \label{Boxplot}
\end{figure}

We excluded all conversation pairs where the generated outputs from either pre-trained or fine-tuned models were non-word, leaving 177 conversation pairs for analysis. We then counted the total number of words included in each response. The mean total number of words per response was 14.05 words ($SD =40.14$). The pre-trained model, on average, generated 75.23 words per response ($SD =114.40$). The fine-tuned model, on average, generated responses that contained 18.44 words ($SD =43.55$). The {\itshape One-way Repeated ANOVA} among the three outputs showed that there was a significant difference ($F (2, 176)= 39.42$, $p<0.001$). {\itshape Tukey’s HSD} paired contrasts \cite{abdi2010tukey} showed that there was a significant difference between the pre-trained model and therapists ($p<0.001$) but not between the fine-tuned model and therapists ($p=0.84$). (For the boxplot, see Fig. \ref{Boxplot})

The example texts below show an example of length differences of the three outputs, where the pre-trained model responded with a longer response than the therapist’s response. Still, the fine-tuned model responded with a length similar to that of the therapist. According to the MACM and conversation theory, \cite{gunaratne2019multi, pask1976conversation}, the conversation context is where there is no new information in the conversation, and the responder should be acknowledging existing information, as the therapist is doing in this example. However, the pre-trained model attempts to add new information with a higher information load, unlike the therapist’s response and the fine-tuned model. This example shows that the fine-tuned model follows the ground truth (i.e., the therapist’s response) much better than the pre-trained model in terms of information load:

\textbf{Input}: {\itshape “You know, it's kind of funny because my brother is four years younger than me and for some reason, I don't remember growing up with him. But in the last seven years taking care of my mom, we got really close. And as far as when we would sit there and talk about our childhood, I say, well where were you during this time, I don't remember you. Because he was home most of the time, and I was outgoing to college and stuff. He started filling me in on stuff going on in the house that I didn't see. And I was shocked. We're starting to talk about things now.”}
\begin{itemize}
\item {\textbf{Therapist’s response}}: {\itshape “That's great.”}
\item {\textbf{The pre-trained model’s response}}: {\itshape “What did you promise your mother? ????? B: I promised her that I would guide and care for Jamy to the best of my knowledge. I don't think she was aware of this until this morning. I love my mother very much and wish her well-loved sons. We both miss her. I cannot afford to receive her well care, but if she needed anything personal, something you wanted, I can follow up or get it. If you need anything from the rest, NOT HARMFUL LIKE 5 TL;DR I will take care of you, and I only ask for pins and drama that you are safe and are safe in my refuge. She is a mental health professional... And most importantly, if you need anything, please do WHAT YOU NEED!!!”}
\item{\textbf{The fine-tuned model’s response}}: {\itshape “Wow!”}
\end{itemize}

\subsection{RQ3 findings: Sentiment analysis}
Table \ref{tab:Descriptive}shows the descriptive statistics of the ten component sentiment scores of the three outputs, of which {\itshape One-way Repeated ANOVA} tested their difference of means.

\begin{table}
  \caption{Descriptive statistics of 10 component sentiment scores of (1) The therapists’ original responses, (2) the pre-trained model’s responses, and (3) the fine-tuned model’s responses.}
  \label{tab:Descriptive}
  \begin{tabular}{p{250pt}p{50pt}p{50pt}p{50pt}}
    \toprule
    Sentiment components ($n=177$) & Therapists’ original responses (Mean, SD) & Pre-trained model’s responses (Mean, SD) & Fine-tuned model’s responses (Mean, SD)\\
    \midrule
    {Negative Adjectives (e.g. “afraid”, “abnormal”)} & -1.04, 0.49 & 0.19, 1.03 & -0.24, 0.80\\
    {Positive Adjectives (e.g. “accessible”, “acceptable”, “accord”)} & 1.72, 1.09 & -0.07, 0.48 & 0.32, 0.89\\
    {Joy (e.g. “tantalizing”, “lovable”, “greeting”)} & 2.82, 3.52 & 0.44, 0.80 & 0.59, 1.56\\
    {Affect Friends and Family (e.g. “bride”, “brother”)} & 0.32, 0.49 & 0.0.23, 0.26 & 0.13, 0.24\\
    {Fear and Disgust (e.g. “inept”, “perverted”)} & 0.02, 0.09 & 0.13, 0.28 & 0.05, 0.20\\
    {Positive Nouns (e.g. “abundance”)} & 0.04, 0.38 & -0.09, 0.50 & 0.02, 0.30\\
    {Respect (e.g. “acknowledgment”, “admiration”)} & 0.02, 0.10 & 0.06, 0.21 & 0.02, 0.12\\
    {Trust Verbs (e.g. “proven”, “merchant”, “pawn”)} & 0.16, 0.23 & 0.13, 0.19 & 0.07, 0.18\\
    {Failure (e.g. “arrest”, “attack”)} & 0.01, 0.07 & 0.05, 0.13 & 0.02, 0.09\\
    {Positive Verbs (e.g., “abound”)} & 0.05, 0.24 & -0.03, 0.44 & 0.04, 0.30\\
    \bottomrule
  \end{tabular}
\end{table}

The {\itshape One-way Repeated ANOVA} among the three outputs showed that “Positive Verbs” ($F(2, 176)=2.88$, $p=0.06$) and “Respect” ($F(2,176)=2.64$, $p=0.07$) did not show significant differences among the three outputs. Both the pre-trained model and fine-tuned model generated similar responses to the therapists on component scores of “Positive Verbs” and “Respect.” The eight component scores showed significantly different results among the three response types ($all p<0.05$). We further tested these eight component scores (i.e., “Negative Adjectives,” “Positive Adjectives,” “Joy,” “Affect Friends and Family,” “Fear and Disgust,” “Positive Nouns,” “Trust Verbs,” “Failure”) with {\itshape Tukey’s HSD} paired contrasts \cite{abdi2010tukey}. We created subgroups by two paired contrast dimensions: the therapists’ original responses versus the pre-trained model outputs and the therapists’ original responses versus the fine-tuned model outputs. Finally, three categories were created because all sentiment components were significantly different in both dimensions. Table \ref{tab:sentimentcomponents} shows the frequency in each category.

\begin{table}[htbp]
  \centering
  \caption{The number of sentiment components whose scores of the pre-trained model or the fine-tuned model significantly differ from therapists’ responses or not: seven out of eight sentiment components differ significantly between the pre-trained model and therapists while five out of eight sentiment components differ significantly between the fine-tuned model and therapists.}
  \begin{tabular}{p{100pt}p{100pt}p{100pt}p{100pt}}
    \toprule
    \multicolumn{2}{c}{{Number of sentiment components}} & \multicolumn{2}{c}{Therapists vs. The pre-trained model} \\
    \cmidrule{3-4}    
    \multicolumn{2}{c}{} & Significantly different & Not significantly different \\
    \midrule
    {Therapists’ vs. The fine-tuned model} & Significantly different & N=4 (Negative Adjectives, Positive Adjectives, Joy, Affect Friends and Family) & N=1 (Trust Verbs) \\
    \cmidrule{2-2}          & Not significantly different & N=3 (Fear and Disgust, Failure, Positive Nouns) & N=0 \\
    \bottomrule
    \end{tabular}%
  \label{tab:sentimentcomponents}%
\end{table}%

\subsubsection{The sentiment component whose score significantly differed between the fine-tuned model and therapists but did not significantly differ between the pre-trained model and therapists.}

Only one sentiment component satisfied this condition: “Trust Verbs” with the component score $p<0.001$. Based on their mean values (mean of “Trust Verbs” of therapists’ original responses= 0.16, mean of “Trust Verbs” of the pre-trained model’s responses = 0.13, mean of “Trust Verbs” of the fine-tuned model’s responses = 0.07), we can see that the fine-tuned model shows fewer positive words on the “Trust Verbs” component compared to that of therapists. An example below demonstrates how the fine-tuned model created fewer positive words on “Trust Verbs” than therapists:

\textbf{Input}: {\itshape “I guess somebody who was actually interested in me and didn't come to the door and say, "How's ?" instead of-- Well, although many people have said, "How are you? How are you holding up?" And many have said, "How can you stand this, day after day?" But that was another affirmation that I am maybe worth something or something like that.”}
\begin{itemize}
\item {\textbf{Therapist’s response}}: {\itshape “Oh, we were glad to be there for you. That's the whole point.”}
\item {\textbf{The pre-trained model’s response}}: {\itshape “How am I holding up NOW?" I: ?"Do I need help him a little? Do I have to? Cause I'm here ?" and any other nonsense. MYMITT HIGH SCHOOL SYMP: ?"MYMITT"<|endoftext|>Background and Synthesis:”}
\item{\textbf{The fine-tuned model’s response}}: {\itshape “Well, you sound like a very open person. I mean, you have a lot of positives around you”}
\end{itemize}

\subsubsection{The sentiment components whose score significantly differed between the pre-trained model and therapists but did not significantly differ between the fine-tuned model and therapists.}
Three components were in this category: “Fear and Disgust” ($p<0.001$), “Failure” ($p<0.001$), and “Positive Nouns” ($p=0.010<0.050$). Based on their means from Table 1, we can see the pre-trained model shows more negative sentiment components, including “Fear and Disgust” and “Failure” and fewer positive sentiment components such as “Positive Nouns” than the therapists. 

For instance, below shows an example of how the “Fear and Disgust” component score of the pre-trained model is more negative than therapists (mean of “Fear and Disgust” of therapists’ original responses= 0.02, mean of “Fear and Disgust” of the pre-trained model’s responses = 0.13, mean of “Fear and Disgust” of the fine-tuned model’s responses = 0.05):

\textbf{Input}: {\itshape “It’s the first year, actually, that they haven’t but, um – so my daughter, actually we are thinking of – it’s her idea – sending her alone on a plane to Florida to visit her grandparents and her cousins.”}
\begin{itemize}
\item {\textbf{Therapist’s response}}: {\itshape “Oh, good. Well, she’s old enough to do that.”}
\item {\textbf{The pre-trained model’s response}}: {\itshape “Albuquerque was‖sport bad, it’s perhaps lucky every food recovered.”}
\item{\textbf{The fine-tuned model’s response}}: {\itshape “? Well, that’s great.”}
\end{itemize}

Below is an example of how the “Failure” component score of the pre-trained model is more negative than therapists (mean of “Failure” of therapists’ original responses= 0.01, mean of “Failure” of the pre-trained model’s responses = 0.05, mean of “Failure” of the fine-tuned model’s responses = 0.02):

\textbf{Input}: {\itshape“So, he's a dear, he's really trying to be helpful too, but it’s just that he is busy with his wife and his family. The only thing is if I do something, I usually have to sandwich it in with everything else that is going on in my life, and I can't just walk away and say, oh, well I'm going to go, we're going to go somewhere new or do something. As you know. So that's why I really appreciated these ladies because they have helped me to just stick up for myself a little bit more.”}
\begin{itemize}
\item {\textbf{Therapist’s response}}: {\itshape “That is great, I'm glad to hear that, and I thank you for sharing it with me. It is helpful to us.”}
\item {\textbf{The pre-trained model’s response}}: {\itshape“? I think the fact that you have said that it is very challenging because you said you were "21 years old me and my marriage, decided at 17 … that it was time to part ways from her" is something you weren't going to do. So far, you've kind of backed off. What has excited you from that?" CSI Yeow Dengpetian Goldman Dear Richard, thank you. Can I return your question about you but point fingers back at it? You said that you wouldn't pursue your law license. I went back and looked at the years I spent at himieva. I found 18 of 18 major events in his name, including pickpocketing, Pimp Involvement, Driver Felony, MS Dump him, Josephine Smith, etc. I really wanted to dig a little deeper on that because at least there was some information there, but now, like Eric, he immediately added insult to injury (...)”}
\item{\textbf{The fine-tuned model’s response}}: {\itshape “Sure, sure.”}
\end{itemize}

Below is an example of how the “Positive Nouns” component score of the pre-trained model is more negative than therapists (mean of “Positive Nouns” of therapists’ original responses= 0.04, mean of “Positive Nouns” of the pre-trained model’s responses = -0.09, mean of “Positive Nouns” of the fine-tuned model’s responses = 0.02):

\textbf{Input}: {\itshape“at many times for you. But there are others in similar circumstances. So hopefully, this will be a little bit of help to you and hopefully not taking up too much of your time 'cause I know things are busy.”}
\begin{itemize}
\item {\textbf{Therapist’s response}}: {\itshape “No, I like it already. I mean, it's great. I mean”}
\item {\textbf{The pre-trained model’s response}}: {\itshape“what is your personal opinion about burritos reppin where people watch tv and look at the [inappropriate sexual slang] bunch?”}
\item{\textbf{The fine-tuned model’s response}}: {\itshape “Well, thank you.”}
\end{itemize}

This example also showed the model using inappropriate slang out of context, which can generate possible harm.

\subsubsection{The sentiment components whose score significantly differed between the pre-trained model and therapists and between the fine-tuned model and therapists.}

Four components were in this category: “Negative Adjectives” (both models: $p<0.001$), “Positive Adjectives” (both models: $p<0.001$), “Joy” (both models: $p<0.001$), “Affect Friends and Family” (the pre-trained model: $p=0.038<0.050$; the fine-tuned model: $p<0.001$). Based on their means from Table 1, we can see both models shows more negative words on components of “Negative Adjectives” and fewer positive words on components of “Positive Adjectives,” “Joy,” and “Affect Friends and Family”  than therapists. 

Below is an example of how both the pre-trained model and the fine-tuned model generated more negative words than therapists on the component of “Negative Adjectives” (mean of “Negative Adjectives” of therapists’ original responses= -1.04, mean of “Negative Adjectives” of the pre-trained model’s responses = 0.19, mean of “Negative Adjectives” of the fine-tuned model’s responses = -0.24):

\textbf{Input}: {\itshape “And even though you have this big yard and you want to go back in your house, you're trapped because they're standing in the whole doorway. And they're making you listen to them, and it's something you disagree with. And you don't want to get in an argument about it. And so finally I had to tell people because there was a little bit of friction going on, and I had to say, you know what? This isn't about you. This whole situation is not about you. It's not about me. It's about our mother. And that's what we have to be happy about and try and deal with. Not our petty problems.”}
\begin{itemize}
\item {\textbf{Therapist’s response}}: {\itshape “That is so great, that is so healthy!”}
\item {\textbf{The pre-trained model’s response}}: {\itshape “But that's so terrible making the ordinary tiny little things feel bigger.”}
\item{\textbf{The fine-tuned model’s response}}: {\itshape “Yeah. I'm sure it was stressful for many people.”}
\end{itemize}
Notice how the therapist focused on staying positive, whereas both GPT-2 models responded negatively.

Similarly, both GPT-2 models--the pre-trained model and the fine-tuned model--generated fewer positive words than therapists on the component of “Positive Adjectives” (mean of “Positive Adjectives” of therapists’ original responses= 1.72, mean of “Positive Adjectives” of the pre-trained model’s responses = -0.07, mean of “Positive Adjectives” of the fine-tuned model’s responses = 0.32):

\textbf{Input}: {\itshape “I know. Can you imagine? Oh, my gosh, I would have just died.”}
\begin{itemize}
\item {\textbf{Therapist’s response}}: {\itshape “Well, it’s a good thing she noticed right away.”}
\item {\textbf{The pre-trained model’s response}}: {\itshape “Wow, I feel you. That timing is incredible for a good old-fashioned real-life rock star like me, earthbound.”}
\item{\textbf{The fine-tuned model’s response}}: {\itshape “I've thought about that. If I ?”}
\end{itemize}

Below is another example of how both GPT-2 models generated fewer positive words than therapists on the component of “Joy” (mean of “Joy” of therapists’ original responses= 2.82, mean of “Joy” of the pre-trained model’s responses = 0.44, mean of “Joy” of the fine-tuned model’s responses = 0.59):

\textbf{Input}: {\itshape “it’s very rewarding. it’s really hard, but it’s, um – I love my job. So –”}
\begin{itemize}
\item {\textbf{Therapist’s response}}: {\itshape “Yeah. Good for you!”}
\item {\textbf{The pre-trained model’s response}}: {\itshape “????NR again. ????”}
\item{\textbf{The fine-tuned model’s response}}: {\itshape “ia, ok.”}
\end{itemize}
The GPT-2 models' responses did not even make sense when the response to this user's input had to be positively reinforced, just like the therapist was attempting to do.

Lastly, the following shows an example of how the GPT-2 models produced less “Affect Friends and Family” compared to the therapists' responses (mean of “Affect Friends and Family” of therapists’ original responses= 0.32, mean of “Affect Friends and Family” of the pre-trained model’s responses = 0.23, mean of “Affect Friends and Family” of the fine-tuned model’s responses = 0.13): 

\textbf{Input}: {\itshape “They are. They are still together, and they are--my folks just turned 80, both of them. And so they are--it's hard to be far from them. But actually, we do a lot of Skyping, a lot of video conferencing together. So that's nice.”}
\begin{itemize}
\item {\textbf{Therapist’s response}}: {\itshape “Well, my folks. But we still have the kids and still, get together. And I know that we're going to leave them--when we go, we're going to leave them in good shape.”}
\item {\textbf{The pre-trained model’s response}}: {\itshape “Hey Glenstar!”}
\item{\textbf{The fine-tuned model’s response}}: {\itshape “What do you do? How does that work?”}
\end{itemize}

In summary, we compared the generative-based model’s pre-trained and fine-tuned models to original therapists’ responses based on three research questions and their corresponding analyses: the proportion of non-word outputs analysis, the length of response (number of words) analysis, and sentiment analysis. Results showed that the fine-tuned model created more non-word outputs than the pre-trained model. For the length of the response analysis, the fine-tuned model performed closer to the therapists compared to the pre-trained model. For the sentiment analysis, both the pre-trained and fine-tuned models were likely to generate more negative or fewer positive outputs than the therapists. Below, we discuss the implications of our findings to designing human-AI interaction in therapy chatbots.

\section{Discussion}
Our findings showed limitations to using generative-based chatbot models for therapy contexts. By assessing the simplified speech quality measures on non-word proportions, length, and sentiment, we saw that much work is still needed in using generative-based language models for therapy contexts, even with its proven technical performance. Especially for health contexts, safety, credibility, personality suitable for context, nuanced responses, etc., are critical for chatbots to adhere to and perform with. Our findings show incredible challenges in designing human-AI interaction, with its unpredictable responses and the need for significantly larger training datasets. Below, we expand on our main findings and discuss potential reasons for the results and what future work can address those challenges.

Both GPT-2 models---pre-trained and fine-tuned---generated a decent portion of non-word outputs. This would confuse users, interfering with the fidelity of patient-therapy chatbot interaction. The reason why both models created non-word outputs and the fine-tuned model created more non-word outputs than the pre-trained model could be the difference of the datasets used for pre-training versus our data used for the fine-tuning and evaluation. The datasets for pre-training are based on the web corpus filtered and modified by humans, and each sentence is a full sentence and well-formatted \cite{radford2019language}. However, the transcripts of therapy conversations for fine-tuning and evaluation were conversation-based dialog pairs compatible with speakers’ habits, rife with informal usages, and partial segments of sentences. Therefore, when models encountered such unfamiliar inputs compared to the data used for pre-training, they might generate non-word outputs accordingly. However, researchers claimed that the OpenAI GPT-2 could process any format of inputs regardless of pre-processing, tokenization, or vocabulary size \cite{radford2019language}, the model still needs improvement. This is a common problem for other pre-trained models, such as BERT \cite{devlin2018bert}, ERNIE (THU) \cite{zhang2019ernie}, BART \cite{lewis2019bart}, RoBERTa \cite{liu2019roberta}, InfoWord \cite{kong2019mutual}, which also uses formal text like Wiki, book, Web, news, and stories \cite{qiu2020pre}. Researchers found a similar phenomenon that BERT is not robust on misspellings \cite{sun2020adv}. To avoid generating non-word outputs, therapy chatbots need to, in real-time, check through all the responses, detect and filter out non-word outputs, and regenerate responses. But such a solution will cause a delay in the model’s responses and cost computational resources. Recent work proved that both generalization and robustness of pre-trained models for natural language processing could be improved through adversarial pre-training or fine-tuning, which uses adversarial examples to train the model so that the model to withstand strong adversarial attacks \cite{liu2020adversarial,zhu2019freelb}. Adversarial training has been widely used in computer vision \cite{shafahi2019adversarial}. However, it is still challenging for text \cite{qiu2020pre}. Future studies should consider using the adversarial training method to reduce the proportion of non-word outputs.

The OpenAI GPT-2 performed well in adjusting the information load of the output with the ground truth (i.e., the therapist’s response) during the fine-tuning process. Therapists in our dataset maintained an ideal information load in their responses based on patients’ input and conversation context. The average length of the responses of the pre-trained model was significantly longer than that of the therapists, which could result in information overload. Researchers found that information overload impacted the speaker’s responsiveness, and the likelihood of response would be suppressed if users were overloaded \cite{gunaratne2020effects}. After fine-tuning, the model generated similar lengths of responses to that of the therapists. This result indicates that the fine-tuning process of GPT-2 potentially adjusted information overload, which is a critical factor in the successful continuation of the conversation. To maintain appropriate information load, language models should decide when to stop generating longer responses. There are trade-offs to generating lengthy adequate textual outputs compared to generating them efficiently in short outputs \cite{gatt2018survey, rieser2009natural}. Early approaches to natural language generation include modular architectures and planning perspectives. Modular architectures treat language generation tasks as the pipeline architecture consisting of sub-tasks, including text plan, sentence plan, and text \cite{reiter1997building}. Planning perspectives view language generation tasks as planning links to satisfy a particular goal, and the generation will stop if the goal is achieved \cite{rieser2009natural}. These approaches tended to sacrifice efficiency to generate short responses in favor of lengthy adequate information \cite{gatt2018survey}. However, from this study, GPT-2 showed that the advanced approach, which emphasizes statistical learning of correspondences between inputs and outputs, can manage the information load through fine-tuning the domain datasets.

The sentiment analysis results imply we must be cautious of directly applying generative-based models without any human filtering in therapy chatbot applications. Both the pre-trained and fine-tuned models were likely to generate more negative adjectives and fewer positive adjectives and words than the therapists. The pre-trained model generated more fear, disgust, and failure sentiments and fewer positive nouns than the therapists, while the fine-tuned model generated fewer trust-related verbs. This phenomenon could cause adverse events when therapy chatbots provide services for therapy contexts. Patients avoid seeking information from the providers if they feel discomfort due to negative responses from the therapist \cite{case2005avoiding, ferlatte2019perceived}. This result may cause potentially harmful interaction, result in ineffective therapy, discourage patients from seeking help when they need mental health support, and result in negative experiences. Patients’ prior experience of seeking mental help greatly impacts the likeliness of seeking mental health help in the future \cite{jon2019perceived, planey2019barriers}. So, developing therapy chatbots that do not have perfectly moderated and approved responses like the approach of this study’s chatbot can be problematic. 

Possible reasons for getting more negative or less positive outputs from models could be from two aspects: the transformer-based model and the dataset size for fine-tuning. The OpenAI GPT-2 is a transformer-based model \cite{oluwatobi2020dlgnet}. The transformer is a model architecture that can directly model the dependency between every two words in a sequence to allow the model to learn language representations and generate outputs like the natural language \cite{qiu2020pre}. Other transformer-based pre-trained models include GPT \cite{radford2018improving}, GPT-3 \cite{brown2020language}, BERT \cite{devlin2018bert}, TransformerXL \cite{dai2019transformer}, ERNIE \cite{zhang2019ernie}, and ERNIE 2.0 \cite{sun2020ernie}. However, the context influencing the direction of conversations is missed in such model architecture because it fails to include human cognitive factors like the speaker’s intents. This phenomenon could result in both models generating more negative or less positive responses than therapists because therapists intend to apply more positive reinforcements in therapy than in ordinary conversations. In addition, the complex, deep non-linear transformer architecture makes it hard to interpret and improve accordingly with a low degree of transparency \cite{qiu2020pre}. The downside of this approach is that we do not have access to understanding the meaning and impact of each parameter of the deep transformer architecture. Explainable artificial intelligence, which aims to make the model architecture transparent and understandable, could be a potential solution to this problem \cite{xu2019explainable, zednik2019solving}. 

In addition, the small size of data for fine-tuning could have influenced the performance of the fine-tuned model. The OpenAI GPT-2 medium model has 345 million parameters, trained over 8 million documents, and 40 GB of text in the pre-training process \cite{radford2019language}. The fine-tuning dataset in this project is less than 7MB, significantly smaller than the data that trained the pre-trained model. Pre-trained models are created to solve this problem. They are expected to avoid overfitting on small data \cite{erhan2010does}, learn universal language representations, and provide a model initialization for better generalization performance \cite{qiu2020pre}. However, this problem still exists due to parameter inefficiency and every application task having its own fine-tuned parameters. Large-scale domain datasets for fine-tuning the pre-trained models are still needed. For medical domains, however, due to privacy concerns, sensitive datasets such as therapist-patient conversations are especially challenging to collect at the level of scale that these models require. Although open-source data platforms in healthcare like Inter-university Consortium for Political and Social Research (ICPSR) \cite{taylor1975inter}, Healthdata.gov (http://www.healthdata.gov), Centers for Disease Control and Prevention data and statistics (https://www.cdc.gov/datastatistics/index.html), CMS Data Navigator (https://dnav.cms.gov/Default.aspx), etc. provide different formats of data including interview, bio-measures, questionnaires, etc., unlike these general datasets, therapy conversation data usually have little chance to be shared in the open-source data platforms due to the confidentiality agreement \cite{bond2002law}. A recent scoping review indicated a delay in applying artificial intelligence for chatbots in mental health compared to chatbots in other fields like customer services \cite{abd2019overview}. Some researchers proposed an improved solution to fix the original parameters of pre-trained models and add small fine-tunable adaption modules for a specific task \cite{stickland2019bert}. Future studies could consider applying such solutions to improve models’ performances. 

The measurements we used to evaluate the chatbot were preliminary. As the next step, we should examine chatbots' responses at a sentence level and a task level to investigate whether the response was suitable as part of a larger context. Now that we have examined response lengths, the next steps are to examine how information overload can be sophisticated for each conversational pair’s context. For instance, the MACM of conversation \cite{gunaratne2019multi} shows how the intentions and acts of the speakers can change the level of information load. Depending on where the conversation is within the therapy context, expectations for information overload should differ depending on these contexts. We should expand sentiment analysis to include further analyzing correct sentiment beyond positive reinforcement-related conversation pairs and investigate the therapy session as a whole. 

\section{Conclusion}
Our study was the first study to evaluate the basic qualities of the meta-information of generative-based therapy chatbot responses. As generative-based models become widely disseminated as AI solutions, and as more healthcare tools adopt AI in the process, we must understand possible opportunities and negative consequences of the impact these new technical solutions will have. Our work contributes to the increasing interest in building therapy chatbots and the rapidly evolving social and everyday computing field. A myriad of AI and machine learning-based solutions become integrated and permeated.


\bibliographystyle{unsrt} 
\bibliography{ms}


\end{document}